\documentclass[12pt]{iopart}

\usepackage{graphicx,iopams}
\begin{document}

\title{Modulation effects on Landau levels in a monolayer
graphene}

\author{J H Ho, Y H Lai, Y H Chiu, and M F Lin}

\address{Department of Physics, National Cheng Kung University,
Tainan, Taiwan 701}
\ead{mflin@mail.ncku.edu.tw}

\begin{abstract}
A monolayer graphene exists in an environment where a uniform
magnetic field interacts a spatially modulated magnetic field. The
spatially modulated magnetic field could affect Landau levels due to
a uniform magnetic field. The modulation effects on Landau levels
are investigated through the Peierl's tight-binding model. The
magneto-electronic properties are dominated by the period, the
strength, and the direction of a spatially modulated magnetic field.
Such a field could induce the growth in dimensionality, the change
of energy dispersions, the destroy of state degeneracy, and the
creation of band-edge states. There are a robust Landau level at
Fermi level and 1D parabolic subbands located around the original
Landau levels, which make density of states exhibit a
delta-function-like structure and many pairs of asymmetric peak
structure, respectively. The density of states and the energies of
band-edge states strongly depend on the strength, but not on the
period and the direction.
\end{abstract}

\maketitle

Recently, the discovery of few-layer graphene\cite{dis:11,dis:12}
has inspired a lot of theoretical and experimental studies in
condensed matter physics and material science. Graphene, as a
nanomaterial, not only has its academic value in condensed matter
physics, but also promises to be the candidate for the
next-generation transistor \cite{dis:11}. Therefore, any advances in
understanding its elementary physical properties, such as electronic
properties
\cite{exp:11,exp:41,theo:11,theo:12,theo:13,theo:14,theo:15,theo:16},
transport properties
\cite{exp:21,exp:22,exp:23,exp:24,theo:21,theo:22,theo:23,theo:24,theo:25}
, optical properties \cite{exp:31,exp:32,theo:31,theo:32,theo:33}
and many body effects \cite{exp:41,theo:41,theo:42} would be very
valuable for finding possible applications. The fact that few-layer
graphene is accessible from experiments means that they can be
controlled in single-atom thick accuracy. From a technical point of
view, this motivates the idea of engineering the electronic
structures \cite{dis:11,exp:11}. The main reason is that the
electronic properties of few-layer graphene strongly depend on the
number of layers and the stacking order \cite{theo:12,theo:14}. One
can further tune its electronic properties by applying a gate
voltage. Such an external field has been shown to induce the
transition of fundamental carriers between electrons and holes in a
monolayer graphene \cite{dis:11}, and lead to opening an energy gap
in a bilayer graphene \cite{exp:11,theo:13}.



The fact that Graphene continuously surprises us should owe to its
exotic electronic structure, where low-energy quasiparicles behave
as massless Dirac particles \cite{exp:21}. In a uniformly
perpendicular magnetic field, the electronic states flock together
and form unusual 0D Landau levels. The level energies, unlike those
of 2D electron gas, are proportional to the square root of the
quantum number and the field strength, rather than being equally
spaced or proportional to the field strength. Recently, these
features have been verified by measurements based on the infrared
spectroscopy \cite{exp:31}. The Landau level quantization reflects
the Dirac nature of its quasiparticle, and is deduced to be
responsible for the observation of unconventional integer quantum
Hall effect \cite{exp:21,exp:23}. On the other hand, the spatially
modulated electric and magnetic fields were predicted to cause
drastic changes in state degeneracy, energy dispersion, band-edge
state, and band width \cite{jon:1,jon:2}. In this work, we would
like to study the modulation effects on Landau levels due to a
spatially modulated magnetic field. The influences of the field
strength, period, and direction on Landau levels will be
investigated in detail through employing Peierl's tight-binding
model.

Graphene is a honeycomb lattice with hexagonal symmetry. The
$\pi$-band structure, formed by $2p_z$ orbital from each carbon
atom, can be calculated through the tight-binding model within the
nearest-neighbor atomic interactions. Since there are two carbon in
the primitive cell, the Hamiltonian can be represented by a $2\times
2$ Hermitian matrix in the space expanded by Bloch functions of two
crystalline sublattices. To apply tight-binding formalism to Bloch
electrons in a magnetic field, we need to consider the extra phase
factor (Peierl's phase), which depends on the vector potential
$\mathbf{A}$, in describing the wave function. Within the
tight-binding scheme\cite{Luttinger:1951}, the wave function in a
magnetic field is expressed as
\begin{eqnarray}
|\Phi_{\mathbf{k}}^n\rangle=\frac{1}{N}\sum_{\mathbf{R}_n}\exp(i\mathbf{k}\cdot\mathbf{R}_n+i\frac{e}{\hbar}G_{\mathbf{R}_n})
|\varphi_{p_z}(\mathbf{r}-\mathbf{R}_n)\rangle.
\end{eqnarray}
$\mathbf{R}_n$ is the lattice vector with sublattice index $n$,
$\varphi_{p_z}$ is the atomic orbital, and
$G_{\mathbf{R}_n}\equiv\int^{\mathbf{r}}_{\mathbf{R}_n}\mathbf{A}(\xi)\cdot
d\xi$ is the phase associated with the magnetic field. It is obvious
that, by disregarding $G_{\mathbf{R}_n}$, the wave function can
recover the original Bloch function. In fact, the phase
$(e/\hbar)G_{\mathbf{R}}$ would lead to the breakdown of the Bloch
condition for the wave function. With this modified Bloch function
$|\Phi_{\mathbf{k}}^n\rangle$, the Hamiltonian matrix element has
already been shown to be that in zero field multiplied by a phase
factor in such a condition that the magnetic field changes slowly as
a function of the lattice constant. It is given by
\begin{eqnarray}
H_{\mathbf{kk'}}^{nn'}&=&\langle\Phi_{\mathbf{k}}^n|H_{\mathbf{B}}|\Phi_{\mathbf{k'}}^{n'}\rangle\nonumber\\
&=&\frac{1}{N}\sum_{\mathbf{R}_n\mathbf{R}_{n'}}\exp\{-i(\mathbf{k\cdot
R}_n-\mathbf{k'\cdot
R}_{n'})-i\Omega(\mathbf{R}_n,\mathbf{R}_{n'})\}H_{\mathbf{R}_n\mathbf{R}_{n'}},
\end{eqnarray}
where
$H_{\mathbf{R}_n\mathbf{R}_{n'}}=\langle\varphi_{p_z}(\mathbf{r}-\mathbf{R}_n)|\mathbf{p}^2/2m+V|\varphi_{p_z}(\mathbf{r}-\mathbf{R}_{n'})\rangle$
is the Hamiltonian matrix element for a single particle in a
periodic lattice potential $V$, and is nonvanishing for
$\mathbf{R}_{n'}=\mathbf{R}_n+\mathbf{a}_n$ ($\mathbf{a}_n$ is the
position vector that connects an atom at $\mathbf{R}_n$ and its
nearest neighbors). The Hamiltonian of Bloch electrons in a magnetic
field is different from that in zero field, in the existence of the
phase difference term
$\Omega(\mathbf{R}_n,\mathbf{R}_{n'})=(e/\hbar)(G_{\mathbf{R}_n}-G_{\mathbf{R}_{n'}})$.
$\Omega(\mathbf{R}_n,\mathbf{R}_{n'})$ could give rise to the
couplings between states with different k's, and hence make solving
band structure more complicated. Generally, the Hamiltonian in
$\mathbf{k}$ space would be irreducible. In other words, the
magnetic field completely destroys the crystal symmetry and
$\mathbf{k}$ is not a good quantum number anymore. However, if
$\Omega(\mathbf{R}_n,\mathbf{R}_{n'})$ is periodic function of
$\mathbf{R}_n$, the periodicity of crystal will be preserved, and
the dimension of the unit cell is in turn determined by the period
of $\Omega$.

The graphene exists in an environment where a uniform perpendicular
magnetic field $B\hat{z}$ interacts a spatially modulated magnetic
field $B'\sin Kx\hat{z}$ along the armchair direction with the
period $l_B=2\pi/K$ [Figure 1]. The magnetic flux, the product of
magnetic field and the hexagonal area, is $\Phi=3\sqrt{3}b^2/2$ in
unit of flux quantum ($\Phi_0=hc/e=4.1356\times 10^{15}[T/m^2]$).
$b=1.42\AA$ is the C-C bond length. The vector potential can be
chosen as $\mathbf{A}=(Bx-B'\cos Kx/K)\hat{y}$. Such a vector
potential would lead to the new periodicity along the armchair
direction. The unit cell is thus enlarged and its dimension is
determined by $\Phi=1/R_B$ and $l_B=3bR_B'$. The parameters $R_B$
and $R_B'$ are both chosen as positive integers to accommodate the
requirement of finite dimensionality of the Hamiltonian matrix, or
finite dimension of a unit cell. The dimensionality is determined by
the least common multiple ($2R_m$) of $2R_B$ and $2R_B'$. Therefore,
the rectangular unit cell contains $4R_m$ carbon atoms and the
Hamiltonian matrix is a $4R_m\times 4R_m$ Hermitian matrix, given by
\begin{eqnarray}
\left(
  \begin{array}{cccccccc}
    0 & q^{\ast} & p^{\ast}_1 & 0 & \ldots & \ldots & 0 & 0 \\
    q & 0 & 0 & p_{2R_m} & 0 & \ldots & \ldots & 0 \\
    p_1 & 0 & 0 & 0 & q^{\ast} & 0 & \ldots & 0 \\
    0 & p^{\ast}_{2R_m} & 0 & 0 & 0 & q & 0 & 0 \\
    \vdots & \ddots & q & 0 & 0 & \ddots & \ddots & 0 \\
    \vdots & \ldots & \ddots & q^{\ast} & \ddots & \ddots & 0 & p_{R_m+1} \\
    0 & \vdots& \vdots & \ddots & \ddots & 0 & \ddots & q \\
    0 & 0 & 0 & 0 & 0 & p^{\ast}_{R_m+1} & q^{\ast} & 0 \\
  \end{array}
\right).
\end{eqnarray}
$p_n\equiv t_{1\mathbf{k}}(n)+t_{2\mathbf{k}}(n)$ and $q\equiv
t_{3\mathbf{k}}$.
$t_{1\mathbf{k}}(n)=\gamma_{0}\exp[\,(ik_{x}b/2+ik_{y}\sqrt{3}b/2)+G_n\,]$,
$t_{2\mathbf{k}}(n)=\gamma_{0}\exp[\,(ik_{x}b/2-ik_{y}\sqrt{3}b/2)-G_n\,]$,
and $t_{3\mathbf{k}}=\gamma_{0}\exp(\,-ik_{x}b\,)$ are three
nearest-neighbor atom-atom interactions, where
$G_n=[\,i\pi\Phi(n-1)+1/6\,]+
[\,-i[6(R_B')^2\Phi'/\pi]\cos[\pi(n-5/6)/R_B']\sin(\pi/6R_B')\,]$ is
the phase caused by the magnetic fields. The magnetic flux
$\Phi'=3\sqrt{3}B' b^2/2$, due to the modulated magnetic field, is
used to characterize its strength. Note that we have adapted special
arrangement of base functions to represent the Hamiltonian matrix as
a band-like matrix. The base functions in the unit cell are chosen
from outside to inside rather than from left to right [Figure 1].
The $\pi$-electronic structure could depend on the direction of the
modulated magnetic field, mainly owing to the anisotropy of
graphene. If the magnetic field is modulated along the zigzag
direction, the corresponding Hamiltonian matrix can be obtained in a
similar way (not shown).

%
%

The unoccupied conduction bands ($E^{c}$'s) are symmetric to the
occupied valence bands ($E^{\nu }$'s) about the Fermi level
$E_{F}=0$. Only the former are discussed in this work. Because the
range of $k_x$ is much smaller than that of $k_y$ for large $R_m$,
it is sufficient just to consider the dispersion along $\hat{k_y}$
in the following discussion. We first discuss the effects due to the
modulated magnetic field along the armchair direction on the low
energy bands resulted from the uniform magnetic field $B=40 T$.
Without $B^{\prime}$, $B$ could make electronic states flock
together and preform Landau levels $E^c(n)$ (n is a nonnegative
integer), as shown in Figure 2(a) by the dotted curves. Such levels
are dispersionless and fourfold degenerate. These well-separated
energy levels suggest that graphene under a uniform magnetic field
could be regarded as a zero-dimensional system. In addition, the
energies of Landau levels obey the simple relation
$E^c(n)\propto\sqrt{nB}$. The perturbed modulated magnetic field
($B'\neq 0$) leads to the drastic changes in energy dispersions,
degeneracies and band-edge states, as shown in Figure 2(a) by the
solid curves at $B'=4 T$ with $R\equiv R_{B'}/R_B=1$. The
dispersionless Landau levels with $n\geq1$ are changed into 1D
parabolic subbands, while the Landau level $E^c(n=0)$ at Fermi level
remains unaltered. Furthermore, such a field destroys the fourfold
degeneracy and creates band-edge states. The magneto-electronic
structure at high- and low-energy regimes are quite similar [Figure
2(b)]. At $B'=0$, the high-energy Landau levels with
$E^c(n)>\gamma_0$ are doubly degenerate, and they are much closer to
each other \cite{jon:3} than the low-energy Landau levels. Under the
field ($B'=40$ T, $R=1$), these Landau levels are totally
transformed into pairs of 1D parabolic subbands. Each pair of
parabolic subbands is nondegenerate and owns two band-edge states at
the zone boundary.

The strength, period, and direction of the modulated magnetic field
strongly affect the low-energy bands, as shown in Figures 2(c)-2(f).
The strength would change the band curvature and cause the shift in
the energies of band-edge states. The energies of local maxima
(minima) increase (decrease) in the increase of $B'$, while the
number ($n_c$) of band-edge states keeps unchanged [Figure 2(c)].
The period presents diverse effects on the properties of the band
structure, as shown in Figure 2(d) with shorter period and Figure
2(e) with longer period compared with Figure 2(a). As the period
varies, the band curvature decreases when $R<1$, whereas it is
unaltered when $R\geq1$. The number of subbands, however, shows
different tendencies in the two situations. When $R\geq1$, the
number of subbands is proportional to $R$, while it is unchanged
when $R<1$. Despite the unchanged number of subbands, more band-edge
states are created with $n_c$ inversely proportional to $R$ when
$R<1$. On the contrary, $n_c$ is unchanged even though there are
more subbands when $R\geq 1$. Furthermore, the period also affects
state degeneracy in a particular manner. The energy bands are
nondegenerate for most $R's$, but they are doubly degenerate as
$1/R$ is an even integer. To see whether the anisotropy is of equal
importance to the $\pi$-electronic structure, the modulated magnetic
field with $(B'=4 T, R=1)$ is applied along the zigzag structure
[Figure 2(f)]. Notice that the actual period in this case is
$\sqrt{3b}R_{B'}$ rather than $3bR_{B'}$ in armchair case. To make
the comparison more precisely, one shall choose a proper $R_{B'}$ of
zigzag case to match the same period. Even with such consideration,
it turns out that the influences of the modulated field along two
definite directions on the Landau levels are similar to each other.

\bigskip Density of states (DOS), which directly reflects the main
features of electronic structures, is defined as
\begin{eqnarray}
D(\omega )=\sum_{\sigma ,h=c,\nu
}\int_{1stBZ}\frac{dk_{x}dk_{y}}{\left( 2\pi \right)
^{2}}\frac{\Gamma }{\pi }\frac{1}{[E^{h}(k_{x},k_{y})-\omega
]^{2}+\Gamma ^{2}}.
\end{eqnarray}
$\Gamma $($=10^{-4}$$\gamma _{0}$) is the phenomenological
broadening parameter. The integration on $k_{x}$, basically, could
be neglected because of the very small range of $k_{x}$. Without
$B'$, $D(\omega)$ is finite at $E_F=0$ and exhibits a lot of
delta-function-like peaks at $\omega\neq0$, as shown by the dashed
curve in Figure 3(a). Such symmetric prominent peaks come from the
0D Landau levels at B=40 T. The distribution of peaks is nonuniform
because of the unequally spaced Landau levels. When a spatially
modulated magnetic field of $(B'=4 T, R=1)$ is applied along
armchair structure, every symmetric peak at $\omega\neq0$ is changed
into a pair of asymmetric prominent peaks except for $n=0$ Landau
level at $E_F=0$, as shown by the solid curve in Figure 3(a). Such
divergent structures come from band-edge states (maxima and minima)
of 1D parabolic subbands. Each pair of asymmetric prominent peaks is
located around the energy of the original Landau level. The
frequencies of the asymmetric prominent peaks are sensitive to the
change of the strength [Figure 3(b)], whereas is insensitive to that
of the period or the direction [Figure 3(c)]. The peak height is
closely related to the band curvatures about band-edge states and
the number of band-edge states, both of which rely on the field
strength and the period as mentioned earlier. The field strength
could decrease the band curvature, so that the peak height gets
weaker when $B'$ increases. On the other hand, the period could
decrease the band curvature and increase the number of band-edge
states, in which the former makes the peak weaker, while the latter
makes the peak stronger. The increasing compensates the decreasing,
therefore, there is no net effect on the peak height. It is thus
deduced that the low-energy spectrum of DOS is only affected by the
strength of the modulated magnetic field even the period has
complicated effects on the Landau levels.

The frequencies of prominent peaks in DOS deserve a closer
investigation. Figure 4(a) shows the relation between the
frequencies ($\omega _{c}$'s) of the first six subpeaks and the
strength at $B=40$ T and $R=1$. The frequency of the first subpeak
is fixed because its robustness against the modulated magnetic field
as mentioned earlier. For other peaks, as the strength increases,
the frequencies increase or decrease depending on whether the
corresponding band-edge states are band maxima or minima. Similarly,
the influence of the period is illustrated by its relation with
$\omega _{c}$'s of the first six subpeaks in Figure 4(b). The
frequencies of those subpeaks are almost unchanged when the period
varies.

In summary, the modulation effects on Landau levels of a monolayer
graphene are investigated through employing the Peierl's
tight-binding model. The magneto-electronic properties are dominated
by the strength, period, and direction of a spatially modulated
magnetic field. Such a field could induce the growth in
dimensionality, the change of energy dispersions, the destroy of
state degeneracy, and the creation of band-edge states. The Landau
levels are transformed into 1D parabolic subbands, but a robust
fourfold degenerate Landau level at Fermi energy is excepted. They
make density of states exhibit many pairs of asymmetric peak
structure and a delta-function-like peak, respectively. Among the
various factors influencing the density of states, only the field
strength is important. The strength, but not the period and
direction, strongly affects the energies of band-edge states, in
which the energies of local maxima (minima) increase (decrease) in
the increase of the field strength. The predicted magneto-electronic
properties could be examined by measurements on the magneto-optical
absorption spectra.

\section*{Acknowledgment}

This work was supported by the National Science Council of Taiwan,
under the Grant Nos. NSC 95-2112-M-006-002.

\newpage

\begin{figure}[p]
\begin{center}\leavevmode
\includegraphics{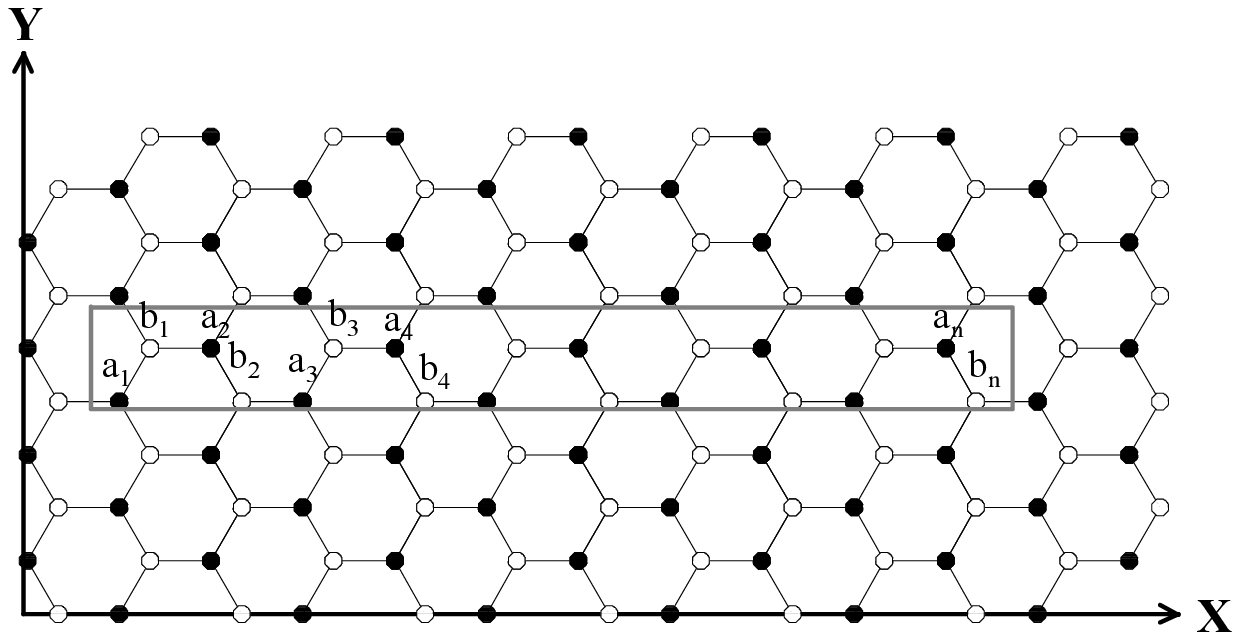}
\caption{The primitive unit cell of a monolayer graphene in a
perpendicular uniform magnetic field and a spatially modulated
magnetic field along the armchair direction.}
\end{center}
\end{figure}

\begin{figure}[p]
\begin{center}\leavevmode
\includegraphics[width=0.8\linewidth]{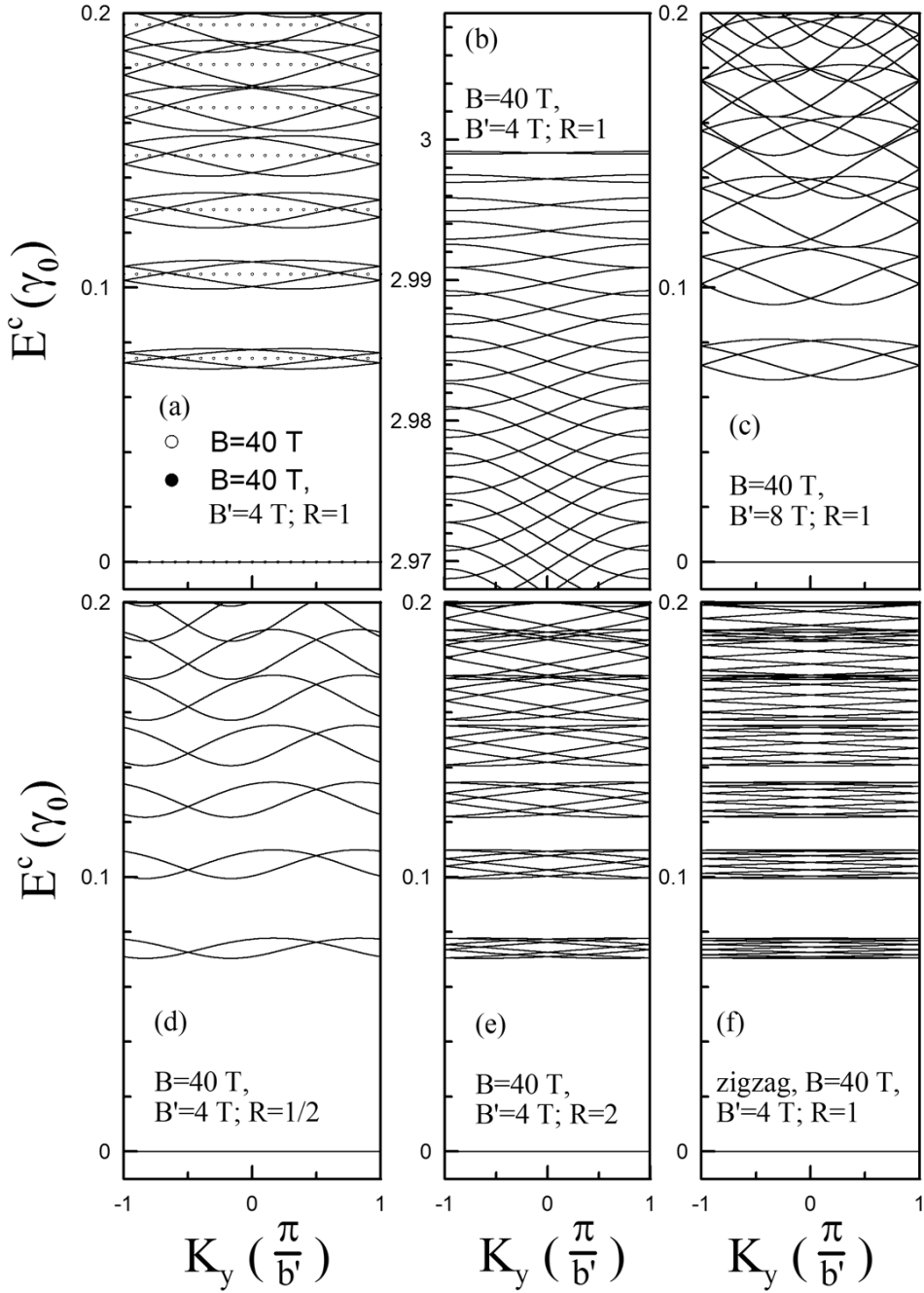}
\caption{The energy bands of (B=40 T, B'=4 T, R=1) at low energy (a)
and high energy (b) regimes. Those of stronger strength B'=8 T,
shorter period R=1/2, longer period R=2, along different direction
(zigzag), respectively, are shown in (c), (d), (e) and (f). The low
energy bands of (B=40 T, B'=0) are also shown in (a) by the dotted
curves. Notice that $b'\equiv\sqrt{3}b\,(3b)$ for armchair (zigzag)
direction.}
\end{center}
\end{figure}

\begin{figure}[p]
\begin{center}\leavevmode
\includegraphics{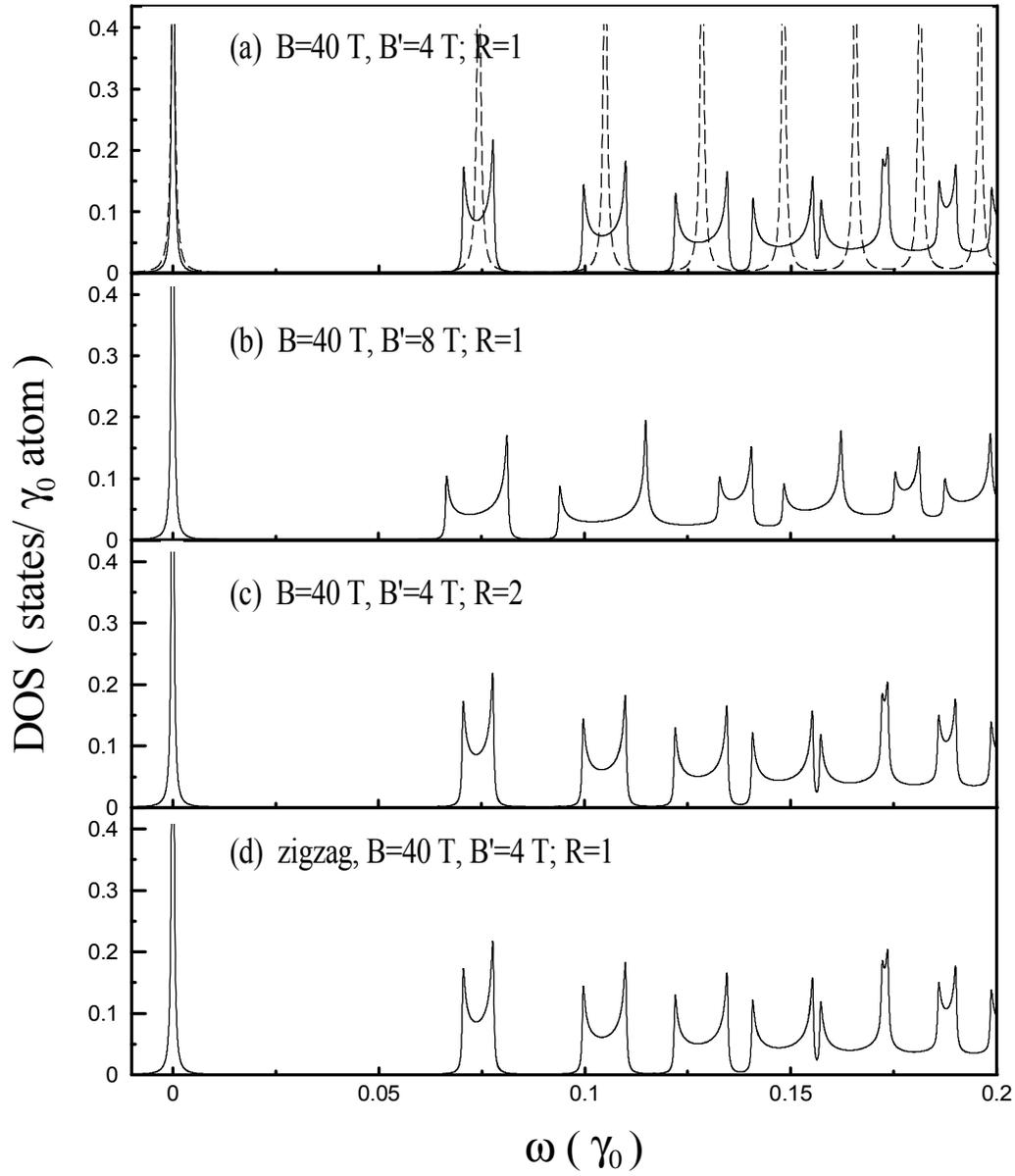}
\caption{The low-frequency density of states along armchair
direction at (a) (B=40 T, B'=4 T, R=1), (b) (B=40 T, B'=8 T, R=1),
(c) (B=40 T, B'=4 T, R=2), and (d) along the zigzag direction at
(B=40 T, B'=4 T, R=1).}
\end{center}
\end{figure}

\begin{figure}[p]
\begin{center}\leavevmode
\includegraphics{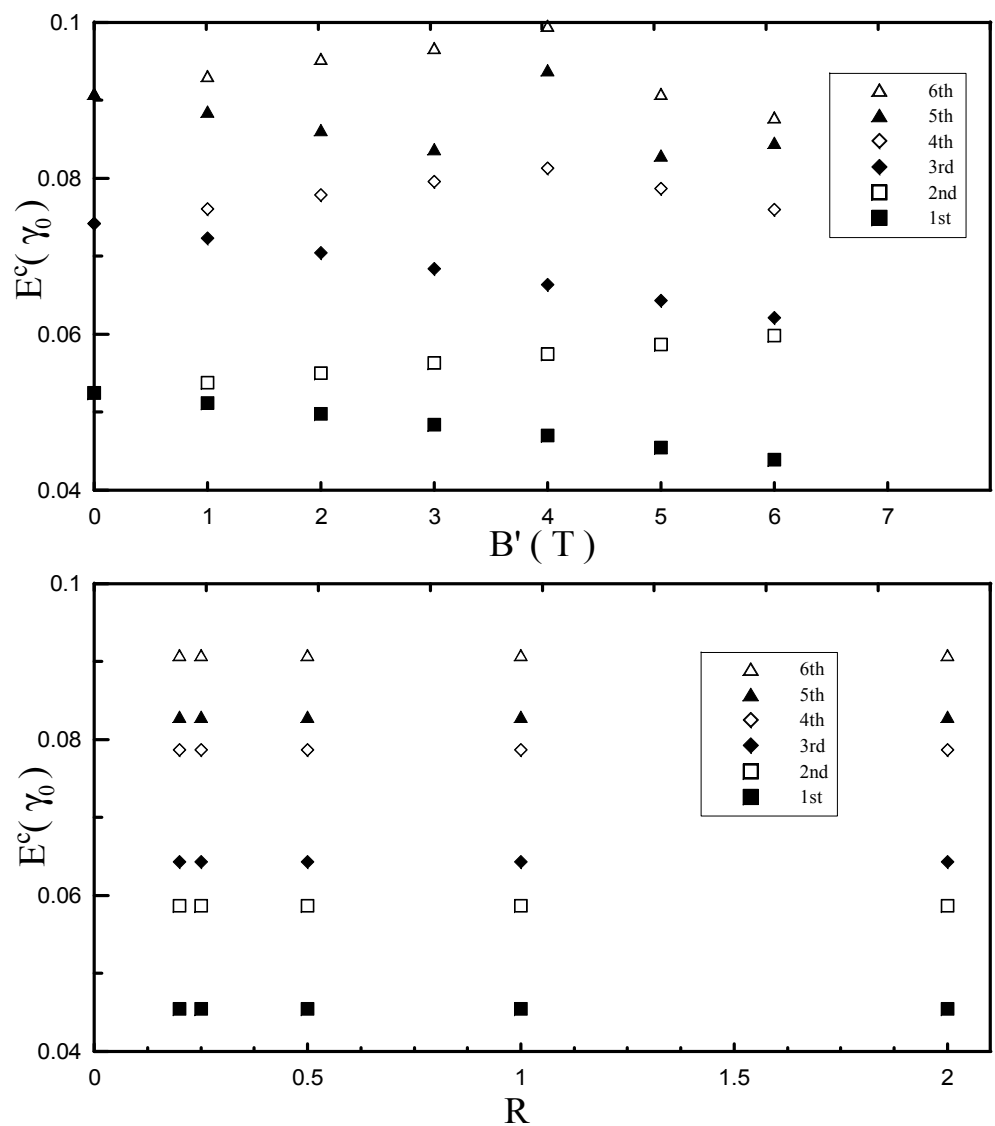}
\caption{Energies ($\omega _{c}$'s) of the first six band-edge
states. (a) and (b) are their dependence on the strength and period
respectively.}
\end{center}
\end{figure}

\newpage

\bibliographystyle{iopart-num}
\bibliography{ref,jon}

\newpage

\end{document}